\documentstyle[12pt]{article}
\def\abstracts#1#2#3{{
        \centering{\begin{minipage}{4.62in}\baselineskip=13pt
        \small
        \centerline{\bf Abstract}
        \vspace*{0.2cm}                % W. Janke (July 1, 1992)
        \parindent=0pt #1\par
        \parindent=18pt #2\par
        \parindent=15pt #3
        \end{minipage} }\par}}
\begin{document}
\vspace*{-4cm}
\addtolength{\textheight}{4cm}
\hfill \parbox{4cm}{ FUB-HEP 18/92 \\
               HLRZ Preprint 76/92} \\
\vspace*{1.0cm}
\centerline{\LARGE \bf High-Temperature Series Analyses of}\\[0.2cm]
\centerline{\LARGE \bf the Classical Heisenberg and XY}\\[0.2cm]
\centerline{\LARGE \bf Model}\\[0.5cm]
\vspace*{0.3cm}
\centerline{\large {\em Joan Adler\/}$^{1,3}$, {\em Christian Holm\/}$^2$ and
                   {\em Wolfhard Janke\/}$^{3,4}$}\\[0.4cm]
\centerline{\large    $^1$ {\small Department of Physics, Technion}}
\centerline{    {\small Haifa 32000, Israel}}\\[0.15cm]
\centerline{\large    $^2$ {\small Institut f\"{u}r Theoretische Physik,
                      Freie Universit\"{a}t Berlin}}
\centerline{    {\small Arnimallee 14, 1000 Berlin 33, Germany}}\\[0.15cm]
\centerline{\large    $^3$ {\small H\"ochstleistungsrechenzentrum,
                      Forschungszentrum J\"ulich}}
\centerline{    {\small Postfach 1913, 5170 J\"ulich, Germany }}\\[0.15cm]
\centerline{\large    $^4$ {\small Institut f\"ur Physik,
                      Johannes Gutenberg-Universit\"at}}
\centerline{    {\small Staudinger Weg 7, Postfach 3980, 6500 Mainz,
                Germany}}\\[1cm]
\vspace*{0.3cm}
\abstracts{}{
Although there is now a good measure of agreement between Monte Carlo
and high-temperature series expansion estimates for Ising  ($n=1$) models,
published results for the critical temperature
from series expansions up to 12{\em th} order
for the three-dimensional
classical Heisenberg ($n=3$) and XY ($n=2$) model do not agree
very well with recent high-precision Monte Carlo estimates.
In order to clarify this discrepancy we have analyzed extended
high-temperature series expansions of the susceptibility, the second
correlation moment, and the second field derivative of the susceptibility,
which have been derived a few years ago by L\"uscher and Weisz
for general $O(n)$ vector spin models on $D$-dimensional hypercubic lattices
up to 14{\em th} order in $K \equiv J/k_B T$.
By analyzing these series expansions in three dimensions
with two different methods that allow for confluent correction terms,
we obtain good agreement with the standard field theory exponent estimates
and with the critical temperature estimates from the new high-precision
 Monte Carlo
simulations. Furthermore, for the Heisenberg model we also reanalyze existing
series for the susceptibility on the BCC lattice up to 11{\em th} order and on
the FCC lattice up to 12{\em th} order using the same methods.
}{}
\thispagestyle{empty}
\addtolength{\textheight}{-4cm}
\newpage
\pagenumbering{arabic}
\section{Introduction}
In the past few years considerable progress has been made in developing
very efficient Monte Carlo (MC) simulation techniques \cite{improved}.
This allows high-precision
computations of the critical coupling and the critical exponents of
continuous phase transitions with an accuracy that is comparable with the
widely accepted estimates derived from field theory \cite{pert,epsilon}.
The third and oldest approach to extract
information about the critical properties of those systems
are analyses of
high-temperature series expansions.
For some standard
models (with notable exceptions including the three-dimensional Ising model
\cite{fisher,adler1} and
certain two-dimensional systems), however,
the critical coupling and the critical exponents calculated by
this method
have much larger error bars and are more vulnerable to systematic errors.
In order to improve this situation
two points are important.
First, more refined methods of analysis than
in the pioneering works must be employed, and second it is obvious that longer
series are needed. The first point  should cause no problem anymore
for continuous phase transitions since over the
years many greatly refined methods have been developed that take into account
various confluent
correction-to-scaling terms and are now available on a routine basis
\cite{gg,acs}. Confluent corrections-to-scaling arise from irrelevant operators
and their neglect can bias critical coupling and critical exponent estimates.
The generation of longer series, however, is still a very demanding
numerical and computational problem, even though it appears to be trivial
in principle.

Significant progress in series generation has been made with star graph
\cite{sc} and
no-free-end (NFE) graph \cite{harris1,harris2,adler2} enumerations
which lead to medium length series in general dimensions
for many systems. However, these approaches are limited by the order of the
existing graph table and not all problems have star or NFE formulations; even
when these exist, the implementation can be quite complex.
For the classical
$O(n)$ vector spin models an important step forward has been made
by L\"uscher and Weisz \cite{lw}, who
applied linked cluster expansion techniques
to compute the expansion
coefficients of the susceptibility, the second correlation moment and
the second field derivative of the susceptibility on $D$-dimensional
hypercubic lattices
up to the 14{\em th} order in the expansion
parameter $K \equiv J/k_B T$ and provided explicit tables for
$1 \le n \le 4$, $2 \le D \le 4$. Moreover,
Butera {\em et al.} \cite{butera} observed that the symmetry of
these models implies
(Schwinger-Dyson) identities between correlation functions that allow a
recursive computation of the series expansion coefficients and reveal their
structure as function of $n$. Combining their result with those of
L\"uscher and Weisz they were able to give the expansion coefficients
in general form as ratios of polynomials in $n$.
Although still one term shorter than the NFE tables \cite{harris1,harris2},
and three terms below
the star graph series of Singh and Chakravarty \cite{sc},
these methods can be used to generate longer series directly,
requiring only larger computer memory and not preexisting graph tables.

The motivation to analyze the extended high-temperature series expansions of
the
Heisenberg model comes from two recent high-precision
MC simulation studies \cite{peczak91,holm92}
of this model on simple cubic (SC) lattices which gave significantly
larger
values for the critical coupling than previous estimates
based on analyses of series expansions up to 12{\em th} order [16-19],
and transfer matrix MC studies \cite{bloete}; see Table~1
(also included is newer MC data \cite{chen93}, that was obtained
after completion of our work).
There are two sources for the
expected improvement. First, on hypercubic lattices two more terms of the
series are known and second, more refined methods taking into account
confluent correction terms are available. For the latter reason we
also reanalyze the long-known but shorter series for the susceptibility
on the body centered cubic (BCC) and face centered cubic (FCC) lattices.
Finally, we present
analyses of the new longer series for the XY ($n=2$) model on the SC lattice.
\section{Model and observables}
We consider the classical $O(n)$ symmetric Heisenberg model with
partition function
\begin{equation}
Z = \prod_i \left[ \int d\Omega_i \right] \exp \left[ K
\sum_{\langle i,j \rangle} \vec{s}_i \cdot \vec{s}_j \right],
\label{eq:1}
\end{equation}
where $K = J/k_B T$ is the reduced inverse temperature,
$\langle i,j \rangle$ denotes nearest-neighbor pairs, and
$\Omega_i$ is the surface of the $n$-dimensional unit sphere
associated with the degrees of freedom of the $n$-dimensional
unit spins $\vec{s}_i$ at each site of a regular three-dimensional lattice.
In this paper we investigate the new longer series for the
Heisenberg ($n=3$) model on a SC lattice, and reanalyze
existing series for the BCC and FCC
lattices. Further we also study the new longer series for the XY ($n=2$) model
on a SC lattice.
In order to estimate the critical couplings and exponents we concentrate
on three observables, the susceptibility
\begin{eqnarray}
\chi &=& \sum_i \langle \vec{s}_0 \cdot \vec{s}_i \rangle
      = \lim_{V \rightarrow \infty} \langle V \left( \frac{1}{V} \sum_i
       \vec{s}_i \right)^2 \rangle \nonumber\\
     &=& A_{\chi} t^{-\gamma}\left[ 1 + a_{\chi} t^{\Delta_1} + b_{\chi} t
 +\dots
        \right], \label{eq:2}
%     &=& \chi_+ t^{-\gamma}\left[ 1 + \chi_1 t^{\Delta_1} + \chi_2 t +\dots
%        \right], \label{eq:2}
\end{eqnarray}
the second correlation moment
\begin{eqnarray}
m^{(2)} &=& \sum_i i^2 \langle \vec{s}_0 \cdot \vec{s}_i \rangle
         =  \chi \frac{\sum_i i^2 \langle \vec{s}_0 \cdot \vec{s}_i \rangle}
                      {\sum_i     \langle \vec{s}_0 \cdot \vec{s}_i \rangle}
\nonumber \\
     &=& A_{m^{(2)}} t^{-(\gamma+2\nu)} \left[ 1 + a_{m^{(2)}} t^{\Delta_1}
         + b_{m^{(2)}} t + \dots  \right], \label{eq:3}
%     &=& m^{(2)}_+ t^{-(\gamma+2\nu)} \left[ 1 + m^{(2)}_1 t^{\Delta_1}
%         + m^{(2)}_2 t + \dots  \right], \label{eq:3}
\end{eqnarray}
and the second field derivative of the susceptibility
\begin{eqnarray}
\chi^{(4)} &=& \frac{3}{n(n+2)} \sum_{i,j,k} \langle \vec{s}_0 \cdot \vec{s}_i
               \,\,\, \vec{s}_j \cdot \vec{s}_k \rangle_c \nonumber \\
     &=& A_{\chi^{(4)}} t^{-(3\gamma+2\beta)} \left[ 1 + a_{\chi^{(4)}}
 t^{\Delta_1}
         + b_{\chi^{(4)}} t +\dots  \right], \label{eq:4}
%     &=& \chi^{(4)}_+ t^{-(3\gamma+2K} \left[ 1 + \chi^{(4)}_1 t^{\Delta_1}
%         + \chi^{(4)}_2 t +\dots  \right], \label{eq:4}
\end{eqnarray}
where $\langle \dots \rangle$ denotes expectation values with respect to the
partition function (\ref{eq:1}) and the subscript $c$ in (\ref{eq:4}) stands
for the connected
part. The second lines in (\ref{eq:2}) -- (\ref{eq:4}) give the assumed
critical behavior where $t \equiv K_c - K > 0$ is the distance
from the critical point
in the high-temperature phase, $\gamma$, $\nu$ and $\beta$ are the
standard critical exponents of the susceptibility, correlation length and
magnetization, respectively, and the terms in square brackets describe the
leading confluent and analytic correction terms. In (\ref{eq:4}) we have
made use of the relation
$\Delta = \gamma + \beta$, where $\Delta$ is the gap exponent. In the
high-temperature phase these observables can be expanded as
\begin{equation}
\chi (n,K) = 1 + \sum_{r=1}^{\infty} a_r(n)K^r,
\label{eq:5}
\end{equation}
\begin{equation}
m^{(2)}(n,K) =  \sum_{r=1}^{\infty} b_r(n)K^r,
\label{eq:6}
\end{equation}
\begin{equation}
\chi^{(4)} (n,K) =  \frac{3}{n(n+2)}
\left[ -2 + \sum_{r=1}^{\infty} d_r(n)K^r \right],
\label{eq:7}
\end{equation}
defining the coefficients $a_r(n)$, $b_r(n)$ and $d_r(n)$
computed in refs.~\cite{lw,butera}. For the convenience of the reader we have
compiled their numerical values for $n=2$ and $n=3$ in Tables~2 and 3.
\section{Methods of analysis}
We analyze the series given in Tables~2 and 3 with two different methods
\cite{klein} that
allow for confluent and analytic correction terms. Taking the susceptibility
as a generic example (and suppressing subscripts) we thus assume a critical
behavior of the form
\begin{equation}
\chi = A t^{-\gamma} \left[ 1 + a t^{\Delta_1} + b t + \dots \right],
\label{eq:8}
\end{equation}
where $\Delta_1 = \nu \omega$ ($\approx 0.55$) is the confluent
correction exponent
and $b t$ a (subleading) analytic correction term. The non-universal
amplitudes $A,a,b$ are assumed to be constant. The $\dots$ inside the
brackets indicate
further higher order corrections of the form $t^{\Delta_m}$, $t^{m+n\Delta_1}$,
which we neglect in our analysis.

In the method referred to as M1, first the leading singularity is
removed by forming
\begin{eqnarray}
B &=& \gamma \chi + t \frac{\partial \chi}{ \partial t} \nonumber \\
  &=& A t^{-\gamma} \left[ \Delta_1 a t^{\Delta_1} + b t + \dots \right].
\label{eq:9}
\end{eqnarray}
Then Pad\'e approximants are applied to the logarithmic derivative of $B$,
\begin{equation}
\frac{\partial \ln B}{\partial t} = \frac{ \Delta_1 (\gamma - \Delta_1) a
t^{\Delta_1-1} + (\gamma - 1) b } { t (\Delta_1 a t^{\Delta_1 -1} + b) },
\label{eq:10}
\end{equation}
yielding for given $K_c$ the confluent correction exponent $\Delta_1$
as function of $\gamma$, $\Delta_1 = \Delta_1(\gamma)$. The optimal set of
values for the parameters $K_c$, $\gamma$ and $\Delta_1$ is determined
visually from the best clustering of different Pad\'e approximants.

In the second method referred to as M2, Pad\'e approximants in a new
variable
\begin{equation}
y = 1 - (1 - K/K_c)^{\Delta_1} = 1 - (t/K_c)^{\Delta_1}
\label{eq:11}
\end{equation}
are applied to
\begin{eqnarray}
t \frac{\partial \ln \chi}{\partial t} &=& -\gamma + \frac{\Delta_1 a
 t^{\Delta_1}
+ b t } { 1 + a t^{\Delta_1} + b t } \nonumber \\
&=& -\gamma - \frac{ \Delta_1 K_c a (y-1) + K_c b (y-1)^{1/\Delta_1}}
{1 - K_c a (y-1) - K_c (y-1)^{1/\Delta_1} },
\label{eq:12}
\end{eqnarray}
yielding for given $K_c$ the exponent $\gamma$ as function of $\Delta_1$,
$\gamma = \gamma(\Delta_1)$. Again the clustering of different Pad\'e
approximants is used to select the optimal set of parameters.

The two methods are complementary and as stressed in App. D of
ref.~\cite{klein} should always
be used in conjunction to avoid spurious results due to so-called
resonances at values of $\Delta_1/n$, $n=2,3,\dots$ in the otherwise more
accurate method M2. The analysis was made with the help of the
recently developed VGS program package \cite{acs} that
makes extensive use of the graphic features of an
X-window environment and allows easy and efficient scanning of the
three-dimensional parameter space.
\section{Results}
% --------------------------------------------------------------
%                       HEISENBERG model
%  -------------------------------------------------------------
\subsection{Heisenberg ($n=3$) model}
\paragraph{SC lattice:} As mentioned in the introduction our main
emphasis was on the Heisenberg model on a SC lattice
since recent high-precision MC simulation studies
\cite{peczak91,holm92} were at
odds with previous high-temperature series expansion analyses [16-19]. In
particular the critical coupling $K_c$ turned out to be significant larger
than widely accepted  series estimates based on expansions up to
12{\em th} order; see Table 1. Our main result from analyses of the longer
14 terms series using methods M1, M2 is that we can
clearly confirm the MC estimates of $K_c$. More precisely for all
three series we get consistent results from methods M1 and M2, and the
three estimates for $K_c$ vary only weakly: $K_c = 0.6928$ from
analyses of $\chi$, $K_c = 0.6930$ from $m^{(2)}$ and $K_c=0.6928$ from
$\chi^{(4)}$. Taking the average of these three values as the final result
we get
\begin{equation}
K_c = 0.6929 \pm 0.0001 \mbox{~~~~\rm (SC lattice)}.
\label{eq:13}
\end{equation}
To illustrate the method of analysis we show for the
susceptibility in Fig.~1 graphs
of the highest near diagonal Pad\'e approximants
to the critical exponent $\gamma$ in the three-parameter space
$K_c$, $\Delta_1$, $\gamma$ computed according to
method M2. A two-dimensional plot of the central slice at
$K_c=0.6928$ is shown in Fig.~2(b). The corresponding plot for
method M1 is displayed in Fig.~2(a). From the point of best clustering
of the different Pad\'e approximants shown in Fig.~2 we read off
\begin{equation}
\gamma = 1.400\pm 0.010,
\label{eq:14}
\end{equation}
and $\Delta_1 = 0.7 \pm 0.2$. Similar analyses of the series for
$m^{(2)}$ yield $\gamma + 2\nu = 2.825 \pm 0.020 $ or inserting
(\ref{eq:14})
\begin{equation}
\nu = 0.712 \pm 0.010,
\label{eq:15}
\end{equation}
and from $\chi^{(4)}$ we get $3\gamma + 2\beta = 4.925 \pm 0.020$ or
using (\ref{eq:14})
\begin{equation}
\beta = 0.363 \pm 0.010.
\label{eq:16}
\end{equation}
Using the scaling relation $\alpha + 2\beta + \gamma = 2$ and the
estimates (\ref{eq:14}),(\ref{eq:16}) we calculate
\begin{equation}
\alpha = -0.125 \pm 0.020.
\label{eq:17}
\end{equation}
Since we have three independent estimates of critical exponents
this result can be used to test the hyperscaling relation $\alpha = 2 - D \nu$.
Using the estimate (\ref{eq:15}) we obtain
\begin{equation}
\alpha = -0.136 \pm 0.030,
\label{eq:18}
\end{equation}
in good agreement with (\ref{eq:17}), thus supporting the hyperscaling
hypothesis. Similarly, the scaling relation $\delta = 1 + \gamma/\beta$
gives
\begin{equation}
\delta = 4.86 \pm 0.10,
\label{eq:18a}
\end{equation}
while the hyperscaling relation $\gamma/\nu = 2 - \eta = D (\delta -1)/
(\delta + 1)$ yields a comparison between
central estimates of
 $\gamma/\nu= 1.966$ from the values
quoted above
and the r.h.s. of the scaling relation
\begin{equation}
   D (\delta -1)/
(\delta + 1)= 1.975,
\label{eq:18b}
\end{equation}
again in good agreement with each other. Our results for the critical
exponents are summarized in Table~4, where they are compared with the
standard field theory values and the results from recent MC
 simulations.
\paragraph{BCC lattice:} The susceptibility series \cite {mckenzie}
consists of only 11 terms, but the overall
behavior is similar. We find optimal convergence at
\begin{equation}
K_c=0.4867\pm0.0001 \mbox{~~~ (BCC lattice),}
\end{equation}
again with $\gamma\approx 1.4$
but with a lower correction-to-scaling exponent
$\Delta_1$ than was seen in the SC case. We quote
central estimates of $\Delta_1 \approx 0.6$ from M1
and $\Delta_1 \approx 0.5$ from M2.
\paragraph{FCC lattice:} For the FCC lattice the
12th order susceptibility series
was analyzed, including corrections-to-scaling,
in ref. \cite{mckenzie}.
It was found that the amplitude of the confluent correction
(with $\Delta_1=0.55$ held fixed at the RG value \cite{pert})
was very small, and that the analytic correction was the dominant one.
We find
\begin{equation}
K_c=0.31475\pm 0.00010 \mbox{~~~ (FCC lattice)}
\end{equation}
and
$\gamma\approx 1.39$, in good agreement with \cite{mckenzie}.
This $\gamma$ is a little lower than our values on the
other lattices, and closer to the values of other calculations.
In contrast to \cite{mckenzie},
we saw clear evidence of a non-analytic
correction-to-scaling at $\Delta_1\approx 0.6$ from the M2 study of a first
derivative  of the susceptibilty series.
%      -------------------------------------------
%                     XY model
%      -------------------------------------------
\subsection{XY ($n=2$) model}
\paragraph{SC lattice:} For the XY model we have only analyzed
the new longer series for the simple cubic lattice. In this case the series
for the susceptibility turned out to be not well-behaved and it was very
difficult to get precise estimates of the critical parameters.
With this
caveat in mind we estimate $K_c = 0.45407$ and $\gamma = 1.325$.
On the other hand the series for $m^{(2)}$ and $\chi^{(4)}$ behaved similar
to the Heisenberg model, i.e., both methods M1 and M2 gave consistent results
and the estimates of $K_c$ from both series agreed with each other,
\begin{equation}
K_c = 0.45414 \pm 0.00007 \mbox{~~~ (SC lattice)}.
\label{eq:19}
\end{equation}
While the previous estimate $K_c=0.4539$ \cite{ferer}
from series analyses is
again lower (and clearly below the error limits of the present study),
our  value (\ref{eq:19}) is consistent with recent Monte
Carlo studies which gave $K_c = 0.45421(8)$ \cite{hase90}
using multiple
and $K_c = 0.4542(1)$ \cite{janke90} using single cluster simulations.
For the exponents we obtain
central estimates of
$\gamma + 2 \nu = 2.67$ from the expansion of
$m^{(2)}$, and  $3\gamma + 2\beta = 4.67$ from
the expansion of $\chi^{(4)}$. The exponents calculated from
these estimates are $\nu = 0.673$,
$\gamma/\nu = 1.970 = 2 - \eta$, and
$\beta = 0.348$. These values are again consistent with
field theoretical estimates \cite{pert,epsilon}.
The scaling relations yield $\alpha =
-0.020$ and $\delta = 4.81$. The hyperscaling relations result in  $\alpha
= -0.018$, and $\gamma/\nu = D (\delta -1)/
(\delta + 1) = 1.968$, again in good agreement with our previous values.
\section{Concluding remarks}
Analyzing new longer series for the Heisenberg ($n=3$) model using
more refined methods than in early works
we obtain for the SC lattice critical parameters that are in
good agreement with completely independent results from two recent MC
simulations. Our reanalyses of existing series for the FCC and BCC lattice
indicate that the improvement comes mainly from the refined methods that
are able to take into account confluent correction terms.
With 14 (or even only 12 or 11) terms these series are, however,
still too short
to compete with the accuracy achieved by field theoretical methods
for critical exponents, or with the precision
claimed from simulations. However, the results
clearly show that there remain no major discrepancies
between series estimates and other calculations. Longer
series clearly stabilize and thus increase the reliability of the estimates
along the lines discussed here, and
it therefore would be very desirable to have a few more terms available.
\section*{Acknowledgements}
We thank the GIF for support of this research. J.A.
thanks H. Herrmann for hospitality at J\"ulich where this project was
commenced, and W.J. gratefully acknowledges a Heisenberg fellowship
by the DFG.
%
%---------------------------------------------------------------
     
%
\newpage
%
%-------------------------------------------------------------------
%                            Tables
%-------------------------------------------------------------------
%
  {\Large\bf Tables}
\vspace*{1cm}
%
%-------------------------------------------------------------------
%                             Table 1
%-------------------------------------------------------------------
%
%
\begin{table}[h]             %[hb]
 \begin{center}
  \begin{tabular}{|l|l|l|}
   \hline
\multicolumn{1}{|c|}{$K_c$} &
\multicolumn{1}{c|}{method} &
\multicolumn{1}{c|}{authors} \\ \hline
 0.692        & 8 terms HTS            & Wood and Rushbrooke (1966) \cite{wood}
   \\
 0.692(4)     & 9 terms HTS            & Joyce and Bowers (1966) \cite{joyce}
\\
 0.6916(2)    & 9 terms HTS            & Ritchie and Fisher  (1972)
 \cite{ritchie}     \\
 0.6924(2)    & 12 terms HTS (Pad\'e)  & McKenzie {\em et al.} (1982)
 \cite{mckenzie} \\
 0.6925(1)    & 12 terms HTS (ratio)   &                                \\
 0.6922(2)    & TMMC ($n \ge 5$)       & Nightingale and Bl\"ote (1988)
 \cite{bloete} \\
 0.6925(3)    & TMMC ($n \ge 6$)       &                                \\
 0.6929(1)    & Metropolis MC          & Peczak {\em et al.} (1991)
 \cite{peczak91}    \\
 0.6930(1)    & 1 Cluster MC           & Holm and Janke (1992) \cite{holm92}
     \\
 0.693035(37) & multiple 1 Cluster MC  & Chen {\em et al.} (1993) \cite{chen93}
     \\
 0.6929(1)    & 14 terms HTS           & this work                      \\
\hline
\end{tabular}
\end{center}
\caption[a]{Estimates of the critical coupling $K_c$ of the
Heisenberg ($n=3$) model on a simple cubic lattice from various sources
(HTS: high-temperature series analysis, TMMC: transfer-matrix Monte Carlo,
MC: Monte Carlo simulation).}
\end{table}
\clearpage
\newpage
%
%-------------------------------------------------------------------
%                             Table 2
%-------------------------------------------------------------------
%
\begin{table}[b]             %[hb]
 \begin{center}
  \begin{tabular}{|c|r|r|r|}
   \hline
\multicolumn{1}{|c|}{order $r$} &
\multicolumn{1}{c|}{$a_r$}  &
\multicolumn{1}{c|}{$b_r$}  &
\multicolumn{1}{c|}{$d_r$} \\ \hline
%
%data......
 1 &           3.0000000000 &           3.0000000000 &         -24.0000000000
\\
 2 &           7.5000000000 &          18.0000000000 &        -160.5000000000
\\
 3 &          18.3750000000 &          72.3750000000 &        -822.0000000000
\\
 4 &          43.5000000000 &         247.5000000000 &       -3576.8125000000
\\
 5 &         102.3437500000 &         770.5937500000 &      -13971.7500000000
\\
 6 &         237.0546875000 &        2261.3437500000 &      -50454.9648437500
\\
 7 &         546.9462890625 &        6360.6650390625 &     -171739.3593750000
\\
 8 &        1252.0048828125 &       17343.7773437500 &     -557978.9429687500
\\
 9 &        2858.8175292969 &       46158.4210449219 &    -1746304.9972656250
\\
10 &        6496.1514078776 &      120515.3193033854 &    -5299323.3505303277
\\
11 &       14735.3746412489 &      309746.4250318739 &   -15671446.8761067708
\\
12 &       33314.7537746853 &      785831.2964274089 &   -45336965.5964835394
\\
13 &       75222.2566392081 &     1971809.9920579093 &  -128702556.1244287884
\\
14 &      169444.4882359232 &     4901417.5916496216 &  -359396456.8541712222
\\
\hline
\end{tabular}
\end{center}
\caption[a]{Expansion coefficients for the XY ($n=2$) model
high-temperature series for a simple cubic lattice.
Given are the expansion coefficients
$a_r$ of the
susceptibility $\chi$, the expansion coefficients
$b_r$ of the second correlation moment $m^{(2)}$,
and the expansion coefficients
$d_r$ of
the second field derivative of the susceptibility $\chi^{(4)}$
up to 14{\em th} order (for
details compare text).}
\end{table}
%}
\clearpage
\newpage
%
%
%-------------------------------------------------------------------
%                             Table 3
%-------------------------------------------------------------------
%

\begin{table}[b]             %[hb]
 \begin{center}
  \begin{tabular}{|c|r|r|r|}
   \hline
\multicolumn{1}{|c|}{order $r$} &
\multicolumn{1}{c|}{$a_r$}  &
\multicolumn{1}{c|}{$b_r$}  &
\multicolumn{1}{c|}{$d_r$} \\ \hline
%
%data......
 1 &           2.0000000000 &           2.0000000000 &         -16.0000000000
\\
 2 &           3.3333333333 &           8.0000000000 &         -71.7333333333
\\
 3 &           5.4222222222 &          21.4222222222 &        -246.0444444444
\\
 4 &           8.5185185185 &          48.7111111111 &        -716.4486772487
\\
 5 &          13.2670194004 &         100.7336860670 &       -1870.2019047619
\\
 6 &          20.3359905938 &         196.1285831864 &       -4508.3329617872
\\
 7 &          30.9989637468 &         365.7050425240 &      -10232.2542817950
\\
 8 &          46.8673402574 &         660.4991803514 &      -22145.7412271162
\\
 9 &          70.6067866595 &        1163.5584276550 &      -46128.4203352476
\\
10 &         105.8320214871 &        2009.6315902889 &      -93088.6148720584
\\
11 &         158.2324753396 &        3414.9732182123 &     -182932.5061463846
\\
12 &         235.7598652836 &        5725.3717946474 &     -351440.3272602895
\\
13 &         350.6189575427 &        9489.5939248535 &     -662121.9818887996
\\
14 &         520.1310140421 &       15575.4527177723 &    -1226410.1925173962
\\
\hline
\end{tabular}
\end{center}
\caption[a]{Expansion coefficients for the classical Heisenberg ($n=3$)
model high-temperature series for the simple cubic lattice. Given are
the expansion coefficients $a_r$ of the
susceptibility $\chi$, the expansion coefficients
$b_r$ of the second correlation moment $m^{(2)}$,
and the expansion coefficients $d_r$ of
the second field derivative of the susceptibility $\chi^{(4)}$
up to 14{\em th} order (for details compare text).}
\end{table}
\clearpage
\newpage
%
%-------------------------------------------------------------------
%                             Table 4
%-------------------------------------------------------------------
%
\begin{table}[b]             %[hb]
 \begin{center}
  \begin{tabular}{|l|l|l|l|l|l|}
   \hline
\multicolumn{1}{|c|}{method}  &
\multicolumn{1}{c|}{$\nu$}    &
\multicolumn{1}{c|}{$\gamma$} &
\multicolumn{1}{c|}{$\beta$}  &
\multicolumn{1}{c|}{$\alpha$} &
\multicolumn{1}{c|}{$\delta$} \\
\hline
$g$-expansion \cite{pert} & 0.705(3) & 1.386(4)  & 0.3645(25) & $-0.115(9)$
  & 4.802(37) \\
$\epsilon$-expansion \cite{epsilon} & 0.710(7) & 1.390(10) & 0.368(4)   &
 $-0.130(21)$ & 4.777(70) \\
MC \cite{peczak91} & 0.706(9) & 1.390(23) & 0.364(7)   &$-0.118(27)$ &
 4.819(36) \\
MC \cite{holm92}   & 0.704(6) & 1.388(14) & 0.362(4)   & $-0.112(18)$ &
 4.837(11) \\
MC \cite{chen93}   & 0.7036(23)&1.3896(70)& 0.3616(31) & $-0.1108(69)$&$-$ \\

this work     & 0.712(10) & 1.400(10)& 0.363(10) & $-0.136(30)$ & 4.86(10)\\
\hline
\end{tabular}
\end{center}
\caption[a]{Critical exponents for the three-dimensional classical
Heisenberg ($n=3$) model from various sources.}
\end{table}
\clearpage
\newpage

%-------------------------------------------------------------------
%                            Figure Headings
%-------------------------------------------------------------------
%
  {\Large\bf Figure Headings}
  \vspace{1in}
  \begin{description}
    \item[\tt\bf Fig. 1:] Graphs of highest near diagonal Pad\'e approximants
    to $\gamma$ in the three-parameter space  $K_c$, $\Delta_1$, $\gamma$
    for method M2. A two-dimensional plot of the central slice at
    $K_c=0.6928$ is shown in Fig.~2(b).
    \item[\tt\bf Fig. 2:] Graphs of highest near diagonal Pad\'e approximants
  to $\gamma$ plotted against $\Delta_1$ at fixed $K_c = 0.6928$ for
  (a) method M1 and (b) method M2.
  \end{description}

\newpage
     \begin{thebibliography}{19}
%---------------------------------------------------------------
%
\bibitem{improved}
For reviews see, e.g.,\\
R.H. Swendsen, J.-S. Wang, and A.M. Ferrenberg, {\em New
Monte Carlo Methods for Improved Efficiency of Computer
Simulations in Statistical Mechanics\/}, in
{\em The Monte Carlo Method in Condensed Matter Physics\/}, edited
by K. Binder (Springer, Berlin, 1991);\\
C.F. Baillie, Int. J. Mod. Phys. {\bf C1}, 91 (1990);\\
A.D. Sokal, {\em Monte Carlo Methods in Statistical
Mechanics: Foundations and New Algorithms\/}, Cours de
Troisi\`eme Cycle de la Physique en Suisse Romande,
(Lausanne, 1989).
%
\bibitem{pert}
J.C. Le Guillou and J. Zinn-Justin, Phys. Rev. Lett. {\bf 39}, 95 (1977);
Phys. Rev. {\bf B21}, 3976 (1980).
%
\bibitem{epsilon}
J.C. Le Guillou and J. Zinn-Justin, J. Physique Lett. {\bf 46}, L137 (1985).
%
\bibitem{fisher}
A.J. Liu and M.E. Fisher, Physica {\bf A156}, 35 (1989).
%
\bibitem{adler1}
J. Adler, J. Phys. {\bf A16}, 3585 (1983).
%
\bibitem{gg}
D. S. Gaunt and A.J. Guttmann, in
{\em Phase Transitions and Critical Phenomena\/},
 Vol.III, edited by C. Domb and M.S. Green,
(Academic Press, New York, 1974), p.181.
%
\bibitem{acs}
J. Adler, I. Chang, and S. Shapira, Technion preprint, Haifa (1993).
%
\bibitem{sc}
R.R.P. Singh  and  S. Chakravarty, Phys. Rev. Lett. {\bf 57},
245 (1986); Phys. Rev. {\bf B36}, 559 (1987).
%
\bibitem{harris1}
A.B. Harris, Phys. Rev. {\bf B26}, 337  (1982).
%
\bibitem{harris2}
A.B. Harris  and  Y. Meir, Phys. Rev. {\bf B36}, 1840 (1987).
%
\bibitem{adler2}
J. Adler, Y. Meir, A. Aharony, A.B. Harris, and L.
Klein, J. Stat. Phys. {\bf 58}, 511 (1990).
%
\bibitem{lw}
M. L\"uscher and P. Weisz, Nucl. Phys. {\bf B300}, 325 (1988).
%
\bibitem{butera}
P. Butera, M. Comi, and G. Marchesini, Phys. Rev. {\bf B41}, 11494 (1990).
%
\bibitem{peczak91}
P. Peczak, A.M. Ferrenberg, and D.P. Landau, Phys. Rev. {\bf B43}, 6087 (1991).
%
\bibitem{holm92}
C. Holm and W. Janke, Phys. Lett. {\bf A173}, 8 (1993),
and preprint FUB-HEP 19/92, HLRZ 77/92, Berlin/J\"ulich (1992).
%
\bibitem{wood}
P.J. Wood and G.S. Rushbrooke, Phys. Rev. Lett. {\bf 17}, 307 (1966).
%
\bibitem{joyce}
G.S. Joyce and R.G. Bowers, Proc. Phys. Soc. {\bf 89}, 776 (1966).
%
\bibitem{ritchie}
D.S. Ritchie and M.E. Fisher, Phys. Rev. {\bf B5}, 2668 (1972).
%
\bibitem{mckenzie}
S. McKenzie, C. Domb, and D.L. Hunter, J. Phys. {\bf A15}, 3899 (1982).
%
\bibitem{bloete}
M.P. Nightingale and H.W.J. Bl\"ote, Phys. Rev. Lett. {\bf 60}, 1562 (1988).
%
\bibitem{chen93}
K. Chen, A.M. Ferrenberg, and D.P. Landau, UGA preprint (1993).
%
\bibitem{klein} L. Klein, J. Adler, A. Aharony, A.B. Harris, and Y. Meir,
Phys. Rev. {\bf B43}, 11249 (1991); and references therein.
%
\bibitem{ferer}
M. Ferer, M.A. Moore, and M. Wortis, Phys. Rev. {\bf B8}, 5205 (1973).
%
\bibitem{hase90}
M. Hasenbusch and S. Meyer, Phys. Lett. {\bf B241}, 238 (1990).
%
\bibitem{janke90}
W. Janke, Phys. Lett. {\bf A148}, 306 (1990).
%
\end{thebibliography}
\end{document}